\documentclass[11pt,sort&compress]{elsarticle}

\input{preamble.tex}

\begin{document}

\hypersetup{
  linkcolor=darkrust,
  citecolor=seagreen,
  urlcolor=darkrust,
  pdfauthor=author,
}

\begin{frontmatter}
    \title{Accelerating Bayesian Optimal Experimental Design via Local Radial Basis Functions: Application to Soft Material Characterization}
    
    \author[add1]{Tianyi Chu\corref{cor1}}
    \ead{tchu72@gatech.edu}
    \author[add2]{Jonathan B.\ Estrada}
    \author[add1,add3,add4]{Spencer H.\ Bryngelson}
 
    \address[add1]{School of Computational Science $\&$ Engineering, Georgia Institute of Technology, Atlanta, GA 30332, USA\vspace{-0.125cm}}
    \address[add2]{Department of Mechanical Engineering, University of Michigan, Ann Arbor, MI 48105, USA\vspace{-0.125cm}}
    \address[add3]{Daniel Guggenheim School of Aerospace Engineering, Georgia Institute of Technology, Atlanta, GA 30332, USA\vspace{-0.125cm}}
    \address[add4]{George W.\ Woodruff School of Mechanical Engineering, Georgia Institute of Technology, Atlanta, GA 30332, USA}
    \cortext[cor1]{Corresponding author}
    \date{}
\end{frontmatter}

\begin{abstract}
We develop a computational approach that significantly improves the efficiency of Bayesian optimal experimental design (BOED) using local radial basis functions (RBFs).
The presented RBF--BOED method uses the intrinsic ability of RBFs to handle scattered parameter points, a property that aligns naturally with the probabilistic sampling inherent in Bayesian methods.
By constructing accurate deterministic surrogates from local neighborhood information, the method enables high-order approximations with reduced computational overhead.
As a result, computing the expected information gain (EIG) requires evaluating only a small uniformly sampled subset of prior parameter values, greatly reducing the number of expensive forward-model simulations needed.
For demonstration, we apply RBF--BOED to optimize a laser-induced cavitation (LIC) experimental setup, where forward simulations follow from inertial microcavitation rheometry (IMR) and characterize the viscoelastic properties of hydrogels.
Two experimental design scenarios, single- and multi-constitutive-model problems, are explored.
Results show that EIG estimates can be obtained at just $8\%$ of the full computational cost in a five-model problem within a two-dimensional design space.
This advance offers a scalable path toward optimal experimental design in soft and biological materials.

\end{abstract}

\blfootnote{Code available at \url{https://github.com/InertialMicrocavitationRheometry/IMR_RBF_BOED}}

\section{Introduction}

Cavitation rheometry has emerged as a powerful technique for characterizing the finite deformation mechanics of soft materials and predicting their behavior under extreme conditions.
Using focused energy sources, such as lasers, ultrasound, or shock waves, this method can induce inertial cavitation within compliant materials, triggering rapid and large deformations at high strain rates ($10^3$--$10^8$\,\unit{\per\second}).
Low-strain-rate techniques ($10^{-3}$--$10^{-1}$\,\unit{\per\second}), including needle- and confinement-induced cavitation, enable quasi-static testing by probing material responses under gradual hydrostatic pressure or mechanical compression~\cite{zimberlin2007cavitation, chockalingam2021probing}. 
Together, these approaches provide complementary insights into mechanical behavior across dynamic and quasi-static regimes.
These capabilities hold promise for biologic and medical applications, including tissue phantom studies, laser surgery diagnosis, and targeted cellular interventions such as DNA manipulation~\citep{mancia2019modeling,vlaisavljevich2016visualizing,brujan2006stress,bailey2003physical,brennen2015cavitation}.

Characterizing realistic soft materials and biotissues, however,
remains challenging under extreme loading rates.
These systems often exhibit high compliance, which complicates stress-strain measurements~\citep{arora1999compliance,chen2010split}, and their nonlinear, time-dependent mechanical behavior deviates markedly from idealized linear elastic models~\citep{lin2009spherical,style2013surface}.
To address these limitations, inertial microcavitation rheometry (IMR) has emerged as a robust method for characterizing soft materials at high strain rates~\citep{estrada2018high}.
By integrating laser-induced cavitation (LIC) experiments with physical bubble dynamics models, IMR enables accurate estimation of viscoelastic properties through time-resolved analysis of bubble radius evolution during cavitation events.
This technique has been applied to characterize widely used biomimetic hydrogels, including
polyacrylamide (PA)~\citep{estrada2018high,yang2020extracting,buyukozturk2022particle},
agarose~\citep{mancia2021acoustic,yang2022mechanical}, collagen~\citep{estrada2021}, fibrin~\citep{abeid2024} and gelatin~\citep{bremer2024ballistic}. 
Recent work by \citet{spratt2021characterizing} further enhanced IMR by integrating data assimilation techniques, providing a scalable framework for bubble-collapse rheometry.
Despite these advances, a bottleneck persists in cavitation experiments: efficiency.
Preparatory procedures---such as chemical treatment, degassing, and swelling protocols---required to create pristine, bubble-free samples impose significant experimental overhead~\citep{lopez2014three,estrada2018high}.
The continuous variability of key material parameters, such as crosslinker density and additive concentration, compounds this challenge.
To address these limitations, an optimal experimental design (OED) strategy accelerates the probing of material responses across diverse physical mechanisms, such as deformation, pressure, and thermal effects.

The goal of the OED approach is to identify the most informative experiments while ensuring robustness against uncertainties---in this case, those inherent to cavitation rheometry.
For this goal, we maximize the expected information gain (EIG), a Bayesian metric estimating the expected reduction in uncertainty about model parameters resulting from a prospective experiment.
EIG is a relative entropy, or Kullback--Leibler (KL) divergence, between the posterior and prior distributions, measuring the information gained from observational data.
By optimizing EIG, the Bayesian OED (BOED) process identifies optimal experimental configurations that maximize the informativeness of measurements while minimizing resource costs.
In addition, BOED enables efficient exploration of high-dimensional parameter spaces in the presence of multiple sources of uncertainty, including variability in experimental configurations and measurement noise.

Our recent work developed a sequential IMR-based BOED approach for characterizing material behavior~\citep{chu2025bayesian}, assuming that the viscoelastic response is governed by two constitutive models: the Neo-Hookean Kelvin--Voigt (NHKV) model~\citep{gaudron2015bubble} and the generalized quadratic law KV (Gen.\ qKV) model~\citep{yang2020extracting,spratt2024numerical}. 
When theoretical models are not available \emph{a priori} to guide the design process, forward simulations across a range of candidate models become necessary.
Prior studies have explored various spherical bubble dynamics models to characterize the viscoelastic properties~\citep{gaudron2015bubble,yang2005model,estrada2018high,wilson2019comparative,barajas2017effects}. 
In such scenarios, the total number of required simulations can grow substantially, making EIG computationally intensive and less likely to converge.
This drawback is not an inherent limitation of OED approaches based on a choice of constitutive model.
Instead, it reflects a broader challenge common to complex modeling scenarios, particularly those involving computationally expensive partial differential equations or other high-fidelity physics-based simulations.

Efforts have been made to reduce the computational cost associated with BOED, particularly in high-dimensional or model-agnostic regimes.
A common strategy involves using EIG estimators with lower sampling complexity.
For example, the sample-reused technique reuses inner-loop Monte Carlo samples across outer loops to reduce cost from $O(N^2)$ to $O(N)$~\citep{huan2013simulation}.
The Monte Carlo with Laplace Approximation (MCLA) estimator~\citep{long2013fast} approximates the posterior distribution as Gaussian, greatly simplifying sampling and lowering computational costs.
By integrating this into the nested Monte Carlo (NMC) framework, the Laplace-based importance sampling NMC estimator achieves higher sampling efficiency while maintaining unbias and avoiding numerical underflow~\citep{beck2018fast}.

Variational inference frameworks like variational OED (VOED)~\citep{foster2019variational} and its normalizing flow extensions~\citep{dong2025variational} amortize inference by approximating posterior or evidence distributions with parametric surrogates. 
These estimators enable the direct evaluation of EIG at candidate designs and are typically followed by design-space exploration using algorithms such as simulated annealing~\citep{muller2005simulation}, interacting
particle systems~\citep{amzal2006bayesian}, stochastic optimization~\citep{huan2013simulation,huan2014gradient,carlon2020nesterov, karimi2021optimal}, and Bayesian optimization (BO)~\citep{foster2019variational,kleinegesse2020bayesian,hase2021gryffin}.
Alternatively, gradient-based optimizations, such as stochastic gradient descent, can circumvent direct EIG evaluation when the EIG gradient is accessible.
Despite these advances, both direct and gradient-based approaches require extensive forward model evaluations to estimate expectations, rendering them computationally prohibitive for complex scenarios that involve expensive simulations.
While existing methods focus on reducing sample complexity, our work addresses a different, and often more critical, bottleneck: the cost of deterministic forward simulations.
We use the radial basis functions (RBFs) to generate large sample sets from a relatively small number of forward simulations, reducing computational cost without sacrificing EIG accuracy.

Rooted in scattered data interpolation, RBF-based methods offer a systematic means of approximating multivariate functions from irregularly distributed data points~\citep{hardy1971multiquadric}. 
While this capability has long been recognized, its potential to address BOED's random sampling demands is largely untapped.
Historically, RBFs were often used as global interpolants over all data points, as reviewed in foundational works~\citep{powell1992theory,powell1999recent,Buhmann_2003}.
Their ability to model spatial correlations, for example, enables effective global optimization strategies~\citep{gutmann2001radial, regis2005constrained, regis2007improved}.
Extensions that incorporate variance estimation further enable the quantification of stochastic system behavior~\citep{regis2007stochastic}.
These properties align closely with Gaussian process (GP) regressions, or Kriging~\citep{rasmussen2003gaussian,williams2006gaussian,oliver1990kriging,kleijnen2009kriging}.
Within these methods, global RBF kernels often serve as stationary covariance functions to construct probabilistic surrogate models that quantify uncertainty for objectives in black-box optimization.
GP regressions form the cornerstone of BO, enabling efficient objective optimization by balancing exploration and exploitation during the search process.
Beyond GP models, the capability of global RBFs to approximate posterior distributions enables their integration into key Bayesian frameworks, including Bayesian model calibration~\citep{bliznyuk2008bayesian}, Bayesian design~\citep{joseph2012bayesian}, and Bayesian inversion~\citep{stuart2018posterior}.
Despite their advantages, the practical application of global RBFs faces some computational challenges.
The computational cost of RBF interpolation grows with dataset size due to the dense matrix operations required for evaluation.
Global RBFs are also prone to ill-conditioning and numerical instability, necessitating regularization techniques such as hyperviscosity~\citep{flyer2012guide, fornberg2011stabilization} and preconditioners~\citep{kansa2000circumventing}.
Yet, such adaptations remain insufficient for BOED, where large-scale sampling and sequential decision-making demand both robustness and efficiency.
To address these limitations, we instead use local RBFs to obtain accurate, efficient, and robust deterministic model surrogates.

The principle behind local RBFs originates from the concept of local Taylor expansions for smooth functions, where the function value at a point can be approximated as a linear combination of nearby values with a desired order of accuracy.
In this context, we assume that the bubble dynamics trajectories evolve smoothly with respect to variations in dimensionless parameters corresponding to the underlying material behavior, such as the Reynolds number and Cauchy number.
By using local RBF stencils, one can effectively capture the behavior of the function within a localized region while avoiding the computational bottlenecks of global interpolation~\citep{shu2003local,tolstykh2000using,wang2002point,wright2003radial}. 
Specifically, polyharmonic splines (PHS) basis functions augmented with polynomials (+poly), introduced by \citet{flyer2016enhancing}, have demonstrated high-order accuracy in solving elliptic partial differential equations (PDEs)~\citep{flyer2016role,bayona2017role,bayona2019insight}.
Unlike traditional RBFs requiring delicate shape parameter tuning~\citep{fasshauer2007choosing,fornberg2004stable}, PHS inherently avoids this issue, simplifying implementation while maintaining robustness.
The inclusion of polynomials ensures consistency with local Taylor expansions~\citep{flyer2016role}, enables far-field decay~\citep{fornberg2002observations}, and supports high-order accuracy~\citep{chu2023rbf,le2023guidelines,shankar2018hyperviscosity}.
Alternative surrogate modeling approaches include projection-based model reduction for deterministic systems~\citep{bui2008model,frangos2010surrogate} and polynomial chaos (PC) expansions for stochastic systems~\citep{xiu2002wiener,ghanem2003stochastic,huan2013simulation}.
Compared to these methods, the local RBF PHS+Poly approach offers adjustable accuracy while efficiently using localized information.
Hence, it enables the approximation of large deterministic sample sets with reduced computational cost.
The concept of local approximation has also been successfully used to improve Markov chain Monte Carlo (MCMC) methods~\citep{conrad2016accelerating,conrad2018parallel}.
Building on these advances, we demonstrate how local RBFs can accelerate the BOED process. 


\begin{figure}[t]
    \centering
    \includegraphics[scale=1]{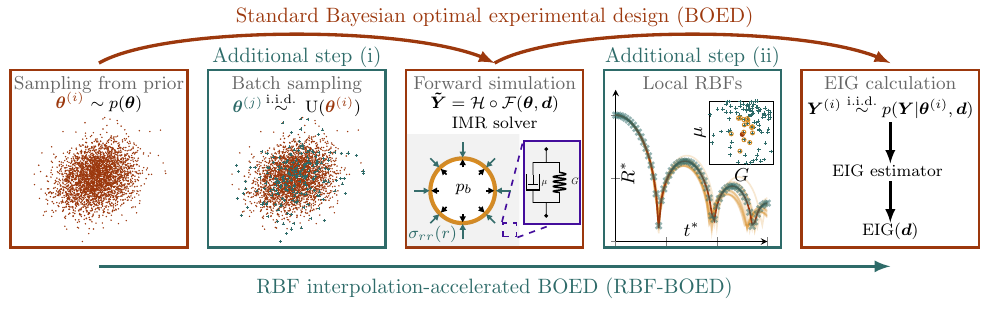}
    \caption{
        Schematic of the RBF--BOED approach.
        RBF--BOED accelerates the standard BOED process through two additional steps:
        (i) evaluate forward models for a small, uniformly sampled batch subset of prior parameters;
        (ii) use local RBF interpolation to construct surrogates, leveraging nearby subsampled points to approximate responses at target parameters.
        The approximated forward model across the entire prior distribution is used for the EIG computation.
        The NHKV model is demonstrated as an example. 
    }
    \label{f:overview}
\end{figure}

\Cref{f:overview} shows an overview of the proposed RBF-accelerated BOED (RBF--BOED) approach.
In \cref{S:IMR}, we introduce the IMR-based BOED, targeting optimal experimental designs in cavitation studies through efficient bubble dynamics simulations.
\Cref{S:RBF-BOED} details the core innovation: a local RBF interpolation scheme to accelerate the BOED process, accompanied by a rigorous error analysis.
The performance of RBF--BOED is demonstrated in \cref{S:results} through two design problems of increasing complexity: single- and multi-constitutive-model LIC scenarios.
\Cref{s:limits,s:conclusions} summarizes the main contributions and limitations.

\section{Objective experiments: laser-induced cavitation (LIC)}\label{S:IMR}

\begin{figure}[t]
    \centering
    \includegraphics[scale=1]{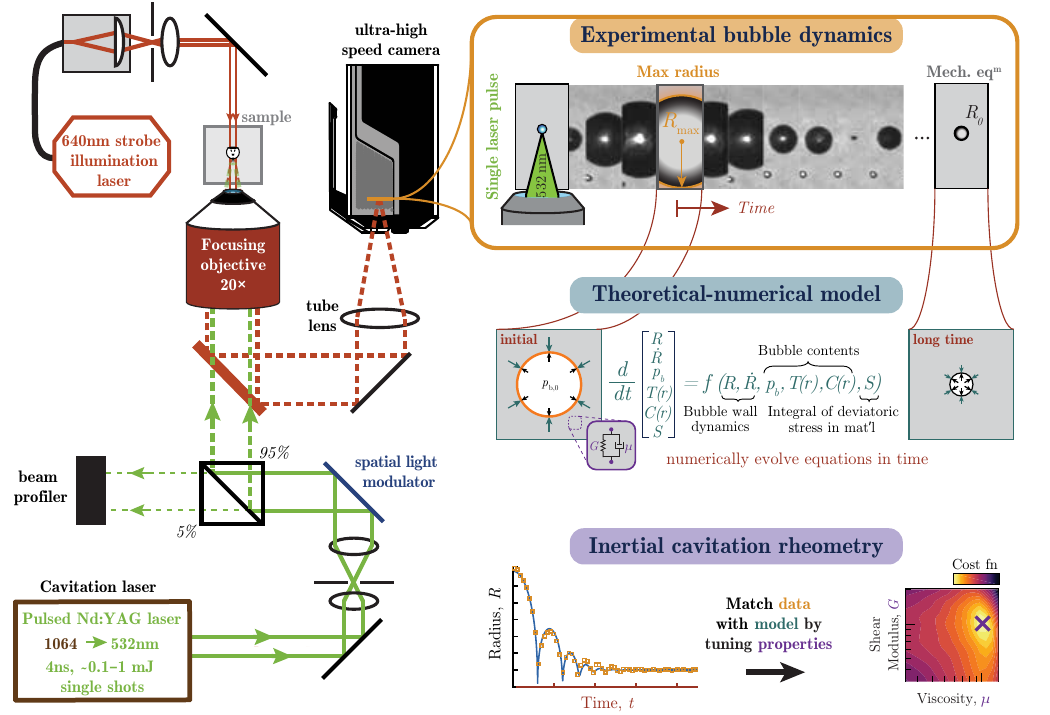}
    \caption{
        Schematic of the LIC experiments and classical inertial microcavitation rheometry (IMR) procedure.
        In a typical laser-induced cavitation (LIC) experiment, a single green laser pulse is reshaped and deposited into a soft material sample via a focusing objective. 
        An ultra-high-speed camera records the event at speeds of $\sim$1 million frames per second; radial dynamics $R(t)$ are acquired via image processing. 
        In IMR, a deterministic theoretical-numerical model is given the experimental maximum and quasi-equilibrium conditions and iteratively matched by adjusting material properties. 
    }
    \label{f:LIC}
\end{figure}

Our objective is to optimize the laser-induced cavitation (LIC) experimental setup for efficiently characterizing the viscoelastic properties of hydrogels by analyzing the bubble radius time history. 
This section presents the inertial microcavitation-based rheometry (IMR) method as a high-strain-rate rheometer, achieved by integrating LIC with physical bubble dynamics models.
\Cref{f:LIC} shows a schematic that illustrates the integration of the LIC experiments with IMR methods for the characterization of soft materials.

\subsection{Bubble dynamics model}

To investigate cavitation in nearly incompressible viscoelastic materials, we use the Keller--Miksis equation~\citep{keller1980bubble} to model the spherically symmetric dynamics of bubble motion.
Derived from mass and momentum conservation, this equation incorporates near-field incompressibility and far-field acoustic radiation, coupled via integration from the bubble wall to infinity.
Nondimensionalized using the maximum bubble radius, $R_{\textrm{max}}$, the far-field pressure, $p_{\infty}$, the surrounding material density $\rho$, and the far-field temperature $T_{\infty}$, the dimensionless Keller--Miksis equation~\citep{keller1980bubble} is given by
\begin{align}
    \left(1-\frac{\dot{R}^*}{c^*}\right)R^*\Ddot{R}^*+\frac{3}{2}\left(1-\frac{\dot{R}^*}{3c^*}\right)\dot{R}^{*^2}=\left(1+\frac{\dot{R}^*}{c^*}+\frac{R^*}{c^*} \dv{}{t^*}\right)\left(p_b^*-\frac{1}{\mathrm{We} \, \dot{R}^*} +S^*-1\right),
\end{align}
which describes the spherical bubble dynamics to first order in the Mach number.
Details of the nondimensionalization are shown in \cref{dimensionless_quantity}. 
The bubble is modeled as a homobaric mixture of water vapor and non-condensable gas, governed by the ideal gas law and characterized by the gas constants $R_v$ and $R_g$, on the time scales of inertial cavitation~\citep{akhatov2001collapse,nigmatulin1981dynamics}.
The time-dependent pressure inside the bubble, $p_b^*(t)$, is coupled to the energy equation~\citep{barajas2017effects,estrada2018high}.
Fick's law and Fourier's law govern the mass and heat transfer of the gases within the bubble.
The LIC model is initialized when the bubble reaches its maximum radius and thermodynamic equilibrium, $R^*(0)=1$.

\begin{table}[ht!]
\centering
\caption{Dimensionless quantities used in this manuscript.}\label{dimensionless_quantity}
{\setlength{\tabcolsep}{9pt}
\begin{tabular}{r l l}
Dimensional & Dimensionless quantity  & Quantity name       \\ \midrule
     &  $U_c=\sqrt{p_{\infty}/\rho}$     & Characteristic velocity  \\
     &  $\lambda = R/R_{\infty}$ & Material stretch ratio   \\
$t$  &  $t^*=t U_c/R_\mathrm{max}$     & Time  \\
$R$  &  $R^*=R/R_\mathrm{max}$ & Bubble-wall radius   \\
$U$  &  $U^*=U/U_c$ & Bubble-wall velocity   \\
$R_{\infty}$  &  $R_{\infty}^*=R_{\infty}/R_\mathrm{max}$ & Equilibrium bubble-wall radius   \\
$c$  &  $c^*=c/U_c$ & Material wave speed   \\
$p_b$  &  $p_b^*=p_b/p_{\infty}$ & Bubble-wall pressure   \\
$p_{\mathrm{v,\,sat}}(T_{\infty})$  & 
$p_{\mathrm{v,\,sat}}^* = p_{\mathrm{v,\,sat}}(T_{\infty})/p_{\infty}$ & Vapor saturation pressure \\ 
$C$  &  $C^*=1/(1+( p_b^*/p^*_{\mathrm{v,\,sat}}-1))R_v/R_g$ &  Vapor concentration   \\
$T$  &  $T^*=T/T_{\infty}$ &  Temperature   \\
$R_\mathrm{max}$  &  $\mathrm{We}=p_{\infty}R_\mathrm{max}/(2\gamma)$ & Weber number   \\
$S$  &  $S^*=S/p_{\infty}$ & Stress integral   \\
$G$  &  $\mathrm{Ca}=p_{\infty}/G$ & Cauchy number   \\
$G_1$  &  $\mathrm{De}=\mu U_c/(G_1 R_{\mathrm{max}})$ & Deborah number   \\
$\mu$  &  $\mathrm{Re}=\rho U_c R_{\mathrm{max}}/\mu$ & Reynolds number   
\end{tabular}
}
\end{table}

This rheometry couples nonlinear bubble dynamics, $R^*(t^*)$, with the time-dependent stress response of the surrounding medium, $S^*(t^*)$.
To parameterize $S^*$ for different gel specimens, appropriate viscoelastic constitutive models are required.
Linear viscoelastic models, such as the linear Kelvin--Voigt (LKV)~\citep{yang2005model} and Maxwell models, are commonly used to describe the evolution of strain rate during bubble expansion and collapse in inertial microcavitation events.
The KV model can be extended to incorporate nonlinear elasticity, such as the Neo-Hookean KV (NHKV) model~\citep{gaudron2015bubble}, which includes a Neo-Hookean elastic term, or the quadratic law KV (qKV) model~\citep{yang2020extracting} and its generalization~\citep{spratt2024numerical}, which include an additional second-order strain-stiffening term.
These extended models capture nonlinear viscoelastic behavior more accurately under high strain rates~\citep{yang2022mechanical}.
The stress integral associated with these models are
\begin{align}
    S^*_{\mathrm{LKV}} &= {-\frac{4 U^*}{\mathrm{Re} R^*}-\frac{4}{3}\overbrace{\left(\frac{1}{\mathrm{Ca}_{\infty}}+\frac{1}{ \mathrm{Ca}_r}\right)}^{1/\mathrm{Ca}}\left[1-\frac{1}{\lambda^3}\right]} \label{eqn:stress integral_LKV},\\
    S^*_{\mathrm{NHKV}} &= {-\frac{4 U^*}{\mathrm{Re} R^*}-\frac{1}{2\mathrm{Ca}}
    \left[5-\frac{4}{\lambda}-\frac{1}{\lambda^4}\right]} \label{eqn:stress integral_NHKV},\\
    S^*_{\mathrm{qKV}} &= {{-\frac{4 U^*}{\mathrm{Re} R^*}-\frac{1}{2\mathrm{Ca}_{\infty}}\left[5-\frac{4}{\lambda}-\frac{1}{\lambda^4}\right]}
    +\frac{\alpha}{\mathrm{Ca}_{\infty}}\left[\frac{177}{20}+\frac{1}{4\lambda^8}+\frac{2}{5\lambda^5}-\frac{3}{2\lambda^4} +\frac{2}{\lambda^2}-\frac{6}{\lambda} -4\lambda \right]}, \label{eqn:stress integral_qKV}
\end{align}
where $\mathrm{Ca}_r$ accounts for the shear modulus contribution from the remaining lumped springs~\citep{estrada2018high}, and
$\alpha$ represents strain stiffening when positive and strain softening when negative~\citep{knowles1977finite}.
To ensure material stability, a lower bound of $\alpha$ is given by $\alpha\geq -2/\left(2\lambda^2+1/\lambda^4-3\right)$.
The NHKV and qKV models correspond to first and second orders of Taylor expansion of the more general Fung model~\citep{fung2013biomechanics}, respectively.
When $\alpha=0$, \cref{eqn:stress integral_qKV} reduces to \cref{eqn:stress integral_NHKV}, in which dynamic shear moduli are used instead of quasistatic moduli to account for the strain stiffening or softening effect during cavitation.
Two generalizations of the LKV and NHKV models---the standard linear solid (SLS)~\citep{zener1949theory} and standard nonlinear solid (SNS)~\citep{estrada2018high} models---further model the time-dependent, non-equilibrium stress response using a logarithmic (Hencky) strain spring:
\begin{align}
    {\mathrm{De}}\dot{S}^*_{\mathrm{SLS}}+ {S}^*_{\mathrm{SLS}} &=  -\frac{4 U^*}{\mathrm{Re} R^*}
    -\frac{4}{3\mathrm{Ca}}\left[1-\frac{1}{\lambda^3}\right]
      -\frac{4 {\mathrm{De}}U^*}{\mathrm{Ca} R^*},\\
     {\mathrm{De}}\dot{S}^*_{\mathrm{SNS}}+ {S}^*_{\mathrm{SNS}}&= 
     -\frac{4 U^*}{\mathrm{Re} R^*}
    -\frac{1}{2\mathrm{Ca}}\left[5-\frac{4}{\lambda}-\frac{1}{\lambda^4}\right]
      -\frac{4 \mathrm{De} U^*}{\mathrm{Ca} R^*(1-\lambda^{-3})}\left[ \frac{3}{14}+\frac{1}{\lambda}-\frac{3}{2\lambda^4}+\frac{2}{7\lambda^7}\right].
\end{align}
These models introduce a Deborah number, $\mathrm{De}$, to characterize the relative timescales of material relaxation and bubble dynamics.
When $\mathrm{De}\to 0$, the SLS and SNS models reduce to the LKV and NHKV models, respectively.
See \citet{estrada2018high} for details on these models.
The constitutive models under consideration are summarized in \cref{database_params}.

\begin{table}[ht!]
\centering
\caption{Summary of constitutive models under consideration.}\label{database_params}
\begin{tabular}{r l l}
Model $\mM$  	 & Description  & Material properties $\vb*{\phi}_{\mM}$   \\\midrule
LKV~\citep{yang2005model} &  Linear Kelvin--Voigt  & $\mathrm{Re},\,\mathrm{Ca}$ \\
NHKV~\citep{gaudron2015bubble} &  Neo-Hookean Kelvin--Voigt  & $\mathrm{Re},\,\mathrm{Ca}$ \\
Gen.\ qKV~\citep{yang2020extracting, spratt2024numerical} & Generalized quadratic Law Kelvin--Voigt    & $\mathrm{Re},\,\mathrm{Ca}_{\infty},\,\alpha$  \\
SLS~\citep{zener1949theory} & Standard Linear Solid  &    $\mathrm{Re},\,\mathrm{Ca},\,\mathrm{De}$\\
SNS~\citep{estrada2018high} & Standard Nonlinear Solid  &    $\mathrm{Re},\,\mathrm{Ca},\,\mathrm{De}$
\end{tabular}
\end{table}

To leverage these theoretical models for experimental design, we construct the modeling parameter within the space ${\Theta}$ by combining the choice of model $\mathcal{M}$ and its corresponding material properties, 
\begin{align}\label{phi_M}
   \vb*{\theta} &\equiv \{\mM,\, \vb*{\phi}_\mM\}.
\end{align}
The probability $p(\vb*{\theta})$ can be separated as the product of the prior probability of the model choice and the conditional probability of the material properties given the model,
\begin{align}
    p(\vb*{\theta}) = p(\mM)p(\vb*{\phi}_{\mM}\vert \mM).
\end{align}
We assume the available models can fully represent the experiments, that is,  $\sum_{\mM} p(\mM) =1$. 

\begin{figure}[ht]
    \centering
    \includegraphics[scale=1]{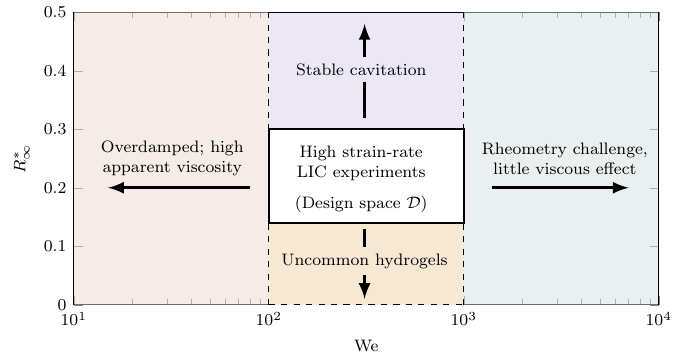}
    \caption{
    Schematic of the design space $\mathcal{D}$ for high strain-rate LIC experiments. 
    }
    \label{f:design space}
\end{figure}

To qualitatively model the LIC experimental configurations, such as laser intensity and equilibrium bubble behavior, we define the design parameter to include the Weber number and equilibrium bubble radius as 
\begin{align}
     \vb*{d} & \equiv \{\text{We},\, R_{\infty}^*\}.
\end{align}
To align with empirical observations, the optimization of these parameters is constrained to experimentally validated ranges $\mathrm{We}\in [100,1000]$ and $R^*_{\infty}\in[0.14,0.3]$~\citep{estrada2018high,yang2020extracting,yang2022mechanical,mancia2021acoustic}.
\Cref{f:design space} shows a schematic of the design space $\mathcal{D}$ for the LIC experiments.

\subsection{Numerical methods}

Given a modeling parameter $\vb*{\theta}$, the forward-time bubble dynamics are simulated as follows.
The state vector consists of the bubble-wall radius, velocity, bubble pressure, stress integral, discretized temperature, and vapor concentration fields inside the bubble:
\begin{align}\label{eq:state}
    \vb*{q}(t) = \{R^*,\,\dot{R}^*,\, p_b^*,\,S^*,\,T(r)^*,\,C(r)^*\}.
\end{align}
Densities, pressures, and temperatures in the surrounding medium are assumed to remain constant, while the bubble’s internal state evolves dynamically.
The discrete-time nonlinear dynamical system is formulated as
\begin{align}\label{IMR_det}
    \vb*{q}_{k+1} = \mF_k(\vb*{q}_k,\vb*{d}), \quad \text{and} \quad
     R^*_{k+1} = \mH(\vb*{q}_{k+1}), 
\end{align}
where $\mF_k$ is the nonlinear operator that governs the system's evolution at each time step, and $\mH$ is the assumed-linear observation function that maps the state $\vb*{q}$ to a point in measurement space. 
\Cref{IMR_det} forms a stiff system due to rapid changes during bubble collapse and rebound.
In practice, a stiff ODE solver with adaptive time-stepping is employed to maintain stability and accuracy, dynamically refining the time step near abrupt radial changes.
Spatial discretization of the temperature and vapor concentration fields introduces high dimensionality, necessitating efficient numerical implementation.

In this study, we designate the bubble radius $R^*$ as the primary observable variable due to its direct measurability in experimental setups.
For a time interval $t\in\left[0,T \right]$ with $N_t$ time steps, the deterministic model outputs, $\btQ = \mqty[ \vb*{q}_1 & \cdots & \vb*{q}_{Nt}]$, and the corresponding bubble dynamics measurements, $\btY = \mqty[ R^*_1 & \cdots & R^*_{Nt}]$, can be collected as
\begin{align}\label{IMR_det_all}
    \btQ = \mF(\vb*{\theta},\vb*{d}) 
    \quad \text{and} \quad
    \btY = \mH(\btQ), 
\end{align}
where $\mF$ is the nonlinear operator that governs the space-time states across all time instances given the initial conditions.
To approximate the actual experimental measurements that span the observation space $\mathcal{Y}$, we incorporate the deterministic IMR solver in \cref{IMR_det_all} with the model error $\vb*{\epsilon}_m$ and the experimental error $\vb*{\epsilon}_e$.
This results in the noisy measurement model
\begin{subequations}\label{IMR_noisy_all}
\begin{align}
    \vb*{Q}_m &= \btQ + \vb*{\epsilon}_m =\mF(\vb*{\theta},\vb*{d}) + \vb*{\epsilon}_m, \quad \text{where} \quad \vb*{\epsilon}_m\sim\mathcal{N}(\vb*{0},\vb*{\Sigma}_m), \quad \text{and} \tag{13a}\\
    \vb*{Y} &  = \vb*{Y}_m + \vb*{\epsilon}_e = \mH(\vb*{Q}_m)+ \vb*{\epsilon}_e , \quad \text{where} \quad \vb*{\epsilon}_e\sim\mathcal{N}(\vb*{0},\vb*{\Sigma}_e). \tag{13b}
\end{align}
\end{subequations}
The observation function $\mH$ is taken to be linear, so \cref{IMR_noisy_all} can be written as 
\begin{align}\label{IMR_noisy_Y}
    \vb*{Y}  = \btY + \vb*{\epsilon}= \mH\circ\mF(\vb*{\theta},\vb*{d})+ \vb*{\epsilon} , \quad \text{where} \quad \vb*{\epsilon}\sim\mathcal{N}(\vb*{0},\vb*{\Sigma}),
\end{align}
where $\vb*{\epsilon}$ is the combined error from the model and the experiments, and the true bubble dynamics, $\btY$, are unobtainable from the measurements.
This procedure follows the method in \citet{freund2015quantitative}.
We now build upon this foundation to develop optimal experimental design (OED) strategies for LIC, described next.


    

\subsection{Bayesian optimal experimental design (BOED)}


The objective of the optimal design procedure is to determine an experiment associated with a design parameter, $\vb*{d}^\star$, that optimally enhances the information about the underlying system compared to prior knowledge.
Mathematically, this is formulated as the maximization of the expected utility function, $u(\vb*{d},\vb*{Y},\vb*{\theta})$, given by
\begin{align}
    \vb*{d}^\star & 
    = \argmax_{\vb*{d}\in\mathcal{D}} \mathbb{E}_{{\Theta},\mathcal{Y}\vert{\vb*{d}}}\{ u(\vb*{d},\vb*{Y},\vb*{\theta}) \}    
    = \argmax_{\vb*{d}\in\mathcal{D}} \int_{\mathcal{Y}}\int_{\Theta} u(\vb*{d},\vb*{Y},\vb*{\theta}) p(\vb*{\theta\vert \vb*{d}},\vb*{Y}) p(\vb*{Y}\vert \vb*{d}) \,\dd\vb*{\theta} \,\dd\vb*{Y}.
\end{align}
Bayes' rule directly relates the prior distribution and the posterior distribution obtained from the prospective experiment via
\begin{align}\label{Bayes}
    \underbrace{p(\vb*{\theta}\vert \vb*{d},\vb*{Y})} _{\mathrm{Posterior}}
 = \frac{ \overbrace{p(\vb*{Y}\vert\vb*{\theta}, \vb*{d})}^{\mathrm{Likelihood}}\,  \overbrace{p(\vb*{\theta}\vert \vb*{d})}^\mathrm{Prior}}{\underbrace{p(\vb*{Y}\vert\vb*{d})}_\mathrm{Evidence}  }.
\end{align}
We assume that $\vb*{d}$ does not provide further information regarding $\vb*{\theta}$ and hence $p(\vb*{\theta}\vert \vb*{d}) = p(\vb*{\theta})$.
A common choice for the utility function is the Kullback--Leibler (KL) divergence between the posterior and prior distributions as a quantification of the relative entropy~\citep{lindley1956measure}.
In this case, the utility function is defined as
\begin{align}
    u(\vb*{d},\vb*{Y},\vb*{\theta}) & 
    = \infdiv{\text{posterior}}{\text{prior}} 
    =  \int_{\Theta} p(\vb*{\theta}\vert \vb*{d},\vb*{Y}) \log{ \left[\frac{p(\vb*{\theta}\vert \vb*{d},\vb*{Y})}{p(\vb*{\theta})}\right]} \,\dd \vb*{\theta} 
    = u(\vb*{d},\vb*{Y}),
\end{align}
which is not a function of the parameters $\vb*{\theta}$.
The expectation of the KL divergence is known as the expected information gain (EIG), defined as
\begin{align}
\mathrm{EIG}(\vb*{d})\equiv  \mathbb{E}_{{\Theta},\mathcal{Y}\vert{\vb*{d}}}\{ u(\vb*{d},\vb*{Y},\vb*{\theta}) \}\label{EIG1}
   &=  
\mathbb{E}_{\Theta} \{\mathbb{E}_{\mathcal{Y}\vert{\Theta,\vb*{d}}}\{  \log{  p(\vb*{\theta}\vert \vb*{d},\vb*{Y}) -\log {p(\vb*{\theta})}}  \}  \} \\
   &=\mathbb{E}_{\Theta} \{\mathbb{E}_{\mathcal{Y}\vert{\Theta,\vb*{d}}}\{  \log{ {p( \vb*{Y}\vert \vb*{\theta}, \vb*{d})}-\log {p(\vb*{Y}\vert\vb*{d})}}  \}  \},\label{EIG2}
\end{align}
where the Bayes' rule in \cref{Bayes} is applied. 
As in~\cref{IMR_noisy_Y}, the likelihood function is modeled as a multivariate Gaussian distribution centered on the ground-truth observable, $\btY$, via
\begin{align}
    p(\vb*{Y}\vert\vb*{\theta}, \vb*{d}) 
    \sim \mathcal{N}(\btY(\vb*{\theta},\vb*{d}),\vb*{\Sigma}),
\end{align}
and the evidence is computed via marginalization over the parameter space
\begin{align}
    p(\vb*{Y}\vert \vb*{d}) &
    = \int_{\Theta} p(\vb*{Y}\vert\vb*{\theta}, \vb*{d}) p(\vb*{\theta}) \,\dd {\vb*{\theta}}.
\end{align}
While \cref{EIG1} or \cref{EIG2} theoretically define the EIG, their direct evaluation is hindered by the computationally intractable double-loop integration over both $\Theta$ and $\mathcal{Y}$.
To address this, BOED relies on numerical estimators to approximate the EIG efficiently.

The most straightforward way to approximate the EIG is using the double-loop Monte Carlo (DLMC) estimator, also known as the nested MC (NMC) estimator~\citep{ryan2003estimating}. 
It is defined as
\begin{align} \label{NMC}
     \mathrm{EIG}(\vb*{d}) 
   & \approx   \mu_{\mathrm{NMC}}(\vb*{d})
   \equiv \frac{1}{N_2} \sum_{j=1}^{N_2}  \log{ \left[\frac{p( \vb*{Y}^{(j)}\vert \vb*{\theta}^{(0,j)}, \vb*{d})}{ \frac{1}{N_1}\sum_{i=1}^{N_1} p(\vb*{Y}^{(j)} \vert \vb*{\theta}^{(i,j)}, \vb*{d})}\right]}, 
\end{align}
where $ \vb*{\theta}^{(i,j)} \overset{\text{i.i.d.}}{\sim} p(\vb*{\theta})$ and $\vb*{Y}^{(j)}  \overset{\text{i.i.d.}}{\sim} p(\vb*{Y}\vert\vb*{\theta}^{(0,j)}, \vb*{d}) $.
The samples $\theta^{(0,j)}$ are used to approximate the outer loop integral,  while $\theta^{(i=1\to N_1,j)}$ are used in the inner loop.
To obtain the dependent pair $(\vb*{\theta}^{(i,j)}, \vb*{Y}^{(i)})$, the importance sampling technique is used: we first draw $\vb*{\theta}^{(i,j)}$ from the prior $p(\vb*{\theta})$, and then draw $\vb*{Y}^{(i)}$ from the conditional distribution $p( \vb*{Y}\vert \vb*{\theta}^{(i,j)}, \vb*{d})$.

Significant efforts have focused on improving the sampling efficiency of estimators for the EIG:
the sample-reused NMC (Re--NMC) estimator~\citep{huan2013simulation}
reduces computational complexity from $O(N_1 N_2)$ to $O(N_2)$ by reusing the inner-loop samples, $\{\vb*{\theta}^{(l)}\}_{l=1}^{N_2}$, across both the outer and inner Monte Carlo loops;
Laplace approximation-based methods~\citep{long2013fast,beck2018fast}
simplify posterior sampling by assuming Gaussianity, significantly lowering computational costs;
Variational approaches like variational optimal experimental design (VOED)~\citep{foster2019variational,foster2020unified,dong2025variational} use parameterized surrogates to approximate posteriors, likelihoods, or evidence.
While existing methods focus on reducing sample complexity, our work targets a different and often more critical bottleneck: the cost of deterministic forward simulations.
Instead of minimizing the number of samples $N_\mathrm{EIG}$ alone, we focus on reducing the total number of forward simulations, denoted by $N_\mathrm{B}$,
to a small batch size sufficient for accurately estimating the EIG.

\section{Radial basis functions (RBF)-based acceleration for BOED}\label{S:RBF-BOED}

For a theoretical model $\mathcal{M}$, the bubble dynamics trajectories evolve smoothly, without discontinuities, for different material parameters. 
In this section, we use this smooth dependence to introduce the radial basis functions (RBF) strategy for accelerating the BOED workflow and validate its accuracy through a rigorous error analysis.

\subsection{Local RBF interpolation}\label{S:RBF}

We aim to approximate the bubble dynamics for a
given target material parameter, $\vb*{\phi}_0$, using a linear combination of $n$ neighboring samples as
\begin{equation}
    \widehat{y}(\vb*{\phi}_0;\vb*{d},t, \mathcal{M}) \,\approx \sum_{j=1}^n w_{j} \tilde y(\vb*{\phi}_j;\vb*{d}, t, \mathcal{M}), \label{Lf} 
\end{equation}
where $\{\vb*{\phi}_j\}_{j=1}^n$ is the set of available samples drawn from the prior distribution and $w_j$ are the unknown interpolation weights.
For simplicity, we omit the dependence on $t$ and $\mathcal{M}$ in the following.
We use the RBF interpolant,
\begin{equation}
    s(\vb*{\phi})=\sum_{j=1}^n \gamma_j \psi(\|\vb*{\phi}-\vb*{\phi}_j\|), \label{rbf_intp}
\end{equation}
to approximate the given function $y(\vb*{\phi})$ by satisfying 
\begin{equation}
    s(\vb*{\phi}_j)=\tilde y(\vb*{\phi}_j), \qquad j= 1,2,\dots,n. \label{rbf_intp_cond}
\end{equation}
Here, $\psi(r)$ is a smooth radial function, $\|\cdot\|$ denotes the standard Euclidean norm, and $\gamma_j$ are the weights associated with the radial basis.
Common choices of RBFs are Gaussians, multiquadrics, and inverse multiquadrics~\citep{fornberg2015solving}.
We use the polyharmonic spline (PHS) class of radial functions, where $\psi(r) =r^m $ if $m$ is odd and $ \psi(r) =r^m\log{r}$ if $m$ is even.  
Readers are referred to recent studies~\citep{bayona2017role,bayona2019role,flyer2016enhancing,flyer2016role} for more details.
Combining \crefrange{Lf}{rbf_intp_cond} leads to the linear system
\begin{equation}
   \underbrace{\mqty[\psi(\|\vb*{\phi}_1-\vb*{\phi}_1\|) & \psi(\|\vb*{\phi}_1-\vb*{\phi}_2\|) & \cdots & \psi(\|\vb*{\phi}_1-\vb*{\phi}_n\|)\\
    \psi(\|\vb*{\phi}_2-\vb*{\phi}_1\|) & \psi(\|\vb*{\phi}_2-\vb*{\phi}_2\|) & \cdots & \psi(\|\vb*{\phi}_2-\vb*{\phi}_n\|)\\
    \vdots & \vdots &  & \vdots\\
       \psi(\|\vb*{\phi}_n-\vb*{\phi}_1\|) & \psi(\|\vb*{\phi}_n-\vb*{\phi}_2\|) & \cdots & \psi(\|\vb*{\phi}_n-\vb*{\phi}_n\|)
    ]}_{\vb*{A}} \mqty[ w_1 \\ w_2 \\ \vdots \\ w_n]
    =\mqty[ \eval{\psi(\|\vb*{\phi}-\vb*{\phi}_1\|)}_{\vb*{\phi}=\vb*{\phi}_0} \\
   \eval{\psi(\|\vb*{\phi}-\vb*{\phi}_2\|)}_{\vb*{\phi}=\vb*{\phi}_0}  \\
   \vdots \\
   \eval{\psi(\|\vb*{\phi}-\vb*{\phi}_n\|)}_{\vb*{\phi}=\vb*{\phi}_0}], \label{w_local}
\end{equation}
which can be solved directly to obtain the weight vector $\vb*{w}=[w_1,\cdots, w_n]^\top$.
Incorporation of polynomial terms (+ Poly) enhances RBF methods by ensuring consistency with local Taylor expansions~\citep{flyer2016enhancing,fornberg2015solving, fornberg2011stabilization,larsson2013stable,wright2006scattered}.
The linear weights in this augmented framework are constrained by moment conditions~\citep{iske2003approximation}, ensuring that the approximation reproduces polynomials up to a specified degree $q$.
These constraints are expressed as
\begin{align}\label{rbf_poly_constrain}
    \sum_{j=1}^n w_j \mathcal{P}_i(\vb*{\phi}_j) =  \mathcal{P}_i(\vb*{\phi}_0) \quad \text{for} \quad 1\leq i\leq  \mqty(q+d_{\vb*{\phi}}\\d_{\vb*{\phi}}),
\end{align}
where $\{\mathcal{P}_i(\vb*{\phi})\}$ denotes the multivariate polynomials up degree to $q$ in the dimensions of $\vb*{\phi}$, denoted as $d_{\vb*{\phi}}$. 
This ensures local polynomial consistency in the RBF approximations~\citep{flyer2016role}, decay in the far-field~\citep{fornberg2002observations}, 
and high order-of-accuracy~\citep{chu2023rbf,chu2024mesh,le2023guidelines,shankar2018hyperviscosity}.
The augmented RBF interpolation becomes
\begin{equation}
   \widehat y(\vb*{\phi}_0) \, \approx\sum_{j=1}^n w_{j} \tilde y(\vb*{\phi}_j) + \sum_{i=1}^{\displaystyle {\tiny \mqty(q+d_{\vb*{\phi}} \\d_{\vb*{\phi}})}} c_i \mathcal{P}_i(\vb*{\phi}_0), \label{Lf_PHS} 
\end{equation}
where $c_i$ are the weights for the corresponding local polynomials. 
A more general and compact representation of \cref{rbf_poly_constrain} and \cref{Lf_PHS} is
\begin{equation} \label{rbf_poly_compact}
   \underbrace{\left[
\begin{array}{c  c }
    \vb*{A} &  \vb*{P} \\ 
   \vb*{P}^{\top} & \vb*{0}
   \end{array}\right]}_{\vb*{A}_\text{aug}}
   \mqty[\vb*{w}\\ \vb*{c}]=\mqty[{\vb*{\psi}}\\
   {\vb*{P}}], 
\end{equation}
where $\vb*{A}$ is the same matrix as defined in \cref{w_local}. 
After solving \cref{rbf_poly_compact}, only the weight vector $\vb*{w}$ is retained for interpolation in \cref{Lf}, while the polynomial coefficients vector, $\vb*{c}$, is omitted.
For a vector-valued target, the approximation is constructed using its neighboring information as 
\begin{equation}\label{rbf_vector}
   \btY(\vb*{\phi}_0) \,\approx \bhY(\vb*{\phi}_0) \equiv \sum_{j=1}^n w_{j} \btY(\vb*{\phi}_j). 
\end{equation}
In practice, a PHS exponent of $m=3$ is often selected due to its favorable trade-off between accuracy and numerical stability~\citep{bayona2019role,flyer2016enhancing,flyer2016role}. 
The local sample size, $n$, depends on the degree of augmented polynomials chosen, $q$, and may vary depending on the distribution of sample points.
\citet{shankar2018hyperviscosity,le2023guidelines} recommended a lower bound on $n$, given by
\begin{align}
    n \geq 2\mqty(q+d_{\vb*{\phi}} \\d_{\vb*{\phi}}) + \lfloor{\ln{\left[2\mqty(q+d_{\vb*{\phi}} \\d_{\vb*{\phi}})\right]}}\rfloor.
\end{align}
The scattered nature of RBF methods naturally aligns with the sampling requirements of Bayesian frameworks. 
Next, we explore their application in BOED.

\subsection{Acceleration for BOED}

We propose the RBF interpolation-accelerated BOED (RBF--BOED) approach to overcome the computational bottlenecks of traditional BOED.
The core innovation lies in approximating the deterministic system response across the parameter space using a scattered batch of forward model evaluations of size $N_\mathrm{B}$.
By using the ability of RBF methods to interpolate scattered data with high accuracy, the RBF--BOED approach enables efficient EIG estimations using orders-of-magnitude fewer forward model evaluations, with $N_\mathrm{B}\ll N_{\mathrm{EIG}}$.

Assuming $N_{\mathrm{EIG}}$ total drawn samples from the prior distribution,
\begin{align}
    \{\vb*{\theta}^{(i)}\}_{i=1}^{N_{\mathrm{EIG}}} \overset{\text{i.i.d.}}{\sim} p(\vb*{\theta})
\end{align}
is required to accurately estimate the EIG.
Each sample requires a forward simulation to evaluate the likelihood, resulting in a total cost of $O(N_{\mathrm{EIG}})$.
Instead of performing $N_{\mathrm{EIG}}$ target forward simulations directly, we uniformly select a subset of $N_{\mathrm{B}}$ samples from the prior samples,
\begin{align}
    \{\vb*{\theta}_{\mathrm{B}}^{(i)}\}_{i=1}^{N_{\mathrm{B}}}\overset{\text{i.i.d.}}{\sim} \mathrm{U}\left(\{\vb*{\theta}^{(i)}\}_{i=1}^{N_{\mathrm{EIG}}}\right),
\end{align}
and perform forward simulations only on this subset.
By construction, the subsampled batch preserves the statistical properties of the prior. 
Each parameter, $\vb*{\theta}^{(i)}$, requires a forward simulation using its associated model $\mathcal{M}$ and model parameter $\vb*{\vb*{\phi}}^{(i)}$.
To approximate this forward simulation, we identify the $n$ nearest neighbors of $\vb*{{\phi}}^{(i)}$ in the batch parameter space, denoted as $\vb*{{\phi}}_{\mathrm{B}}^{(i_1)},\cdots,\vb*{{\phi}}_{\mathrm{B}}^{(i_n)}$.
The corresponding precomputed evaluations,
$\btY_{\mathrm{B}}^{(i_1)},\cdots,\btY_{\mathrm{B}}^{(i_n)}$, 
are then used to approximate the target forward model through \cref{rbf_vector}, 
\begin{align}
    \btY^{(i)}\approx \bhY^{(i)} = \sum_{j=1}^n w_{j}^{(i_j)} \btY(\vb*{\phi}_\mathrm{B}^{(i_j)}),
\end{align}
 This process is repeated for each model $\mathcal{M}$ under consideration. 
The obtained approximated evaluations can be directly used for the EIG estimation without the need for additional forward simulations.
An algorithm for RBF--BOED is outlined in \cref{alg:BOED_RBF}. 
This procedure identifies the optimal prospective experiment over $N_\mathrm{I}$ iterations of the design search.

\begin{figure}
    \centering
    \includegraphics[scale=1]{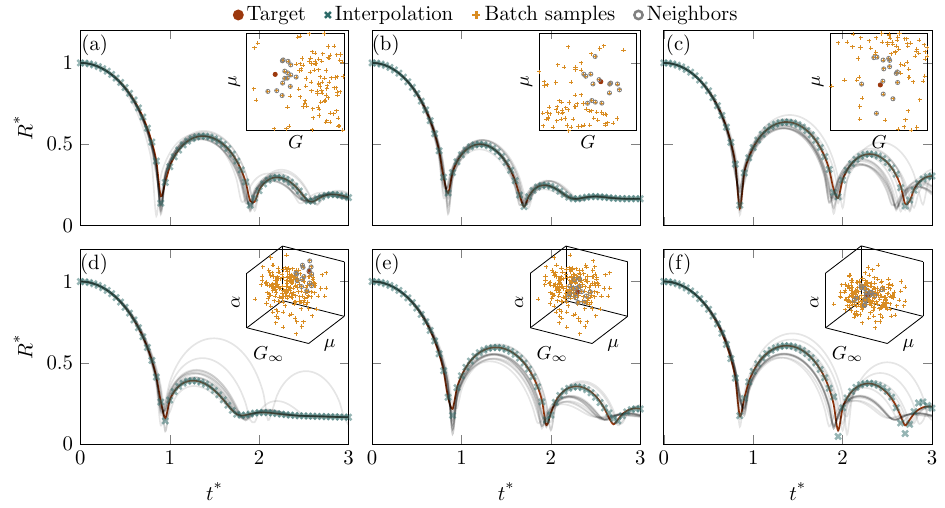}
    \caption{
        Interpolation phase of RBF--BOED for LIC experiments via: (a,b,c) NHKV; (d,e,f) Gen.\ qKV.
    }
    \label{f:IMR_interpolation}
\end{figure}

\Cref{f:IMR_interpolation} exemplifies the interpolation phase of RBF--BOED for LIC experiments using the NHKV and Gen.\ qKV models, which involve $2$ and $3$ material properties, respectively.
The strategic selection of neighboring batch samples ensures simulations exhibit consistent physical trends with moderate variations, enabling localized exploitation of the parameter space while maintaining computational efficiency.
Compared to other stages of bubble dynamics, interpolation errors are higher during bubble collapse, a phase characterized by inherently greater measurement uncertainties.
Overall, RBF interpolations achieve high-fidelity approximations of the target simulations, demonstrating robustness across the tested cases.

We highlight that the RBF--BOED approach aims to reduce the number of forward simulations to $N_{\mathrm{B}}$, 
while maintaining the full sample size $N_{\mathrm{EIG}}$, necessary for accurate EIG estimation. 
Consequently, this method can be combined with other established techniques, such as Re--NMC and VOED, that aim to reduce the required number of samples.
In practice, we adopt Re--NMC to further reduce the required EIG sample size, $N_{\mathrm{EIG}}$.
This method eases the application of various optimization strategies within the design space by reducing the computational cost of each EIG evaluation.
Bayesian optimization (BO) is one of the most widely adopted strategies for design-space exploration~\citep{foster2019variational,kleinegesse2020bayesian,hase2021gryffin}.
We refer the reader to \citet{shahriari2015taking,snoek2012practical} for a comprehensive review and practical implementation of BO. 
Central to BO is a Gaussian Process (GP) regression model, which approximates the EIG landscape using available evaluations and global RBF kernels.
To capture spatial correlations across the design space, we adopt the ARD Matérn 5/2 kernel,
\begin{align}\label{M52}
        \psi_\mathrm{M52}(r)=\eta_0\left(1+\sqrt{5}r+\frac{3}{2}r^2\right)\mathrm{exp}\left(-\sqrt{5}r\right),
\end{align}
where $r$ is the normalized distance between the points in design space, and $\eta_0$ is the covariance amplitude~\citep{snoek2012practical}.
This kernel is well-suited for representing smooth and moderately varying functions, and its selection reflects the empirically observed continuity of the material response within the target design space.
In practice, we construct the GP regression model using the \texttt{MATLAB} function \texttt{fitrgp}, which takes as input the available EIG evaluations and the kernel function in \cref{M52}.
By enabling the \texttt{OptimizeHyperparameters} option, the function minimizes the cross-validation loss through hyperparameter tuning, such as kernel parameters and standardization. 
Using the predictive mean and variance functions, the expected improvement criterion is then applied as the acquisition function to explore the design space.


\begin{algorithm}[ht!]
    \caption{
        RBF-accelerated Bayesian optimal experimental design (refer to \cref{f:overview} for graphical illustration) 
    }\label{alg:BOED_RBF}
    \hspace*{\algorithmicindent} \textbf{Input:} prior $p(\vb*{\theta})$, likelihood function, EIG sample size $N_\mathrm{EIG}$, batch sample size $N_{B}$, iterations of searching $N_{\mathrm{I}}$, RBF parameters $(n,m,q)$ \\
    \hspace*{\algorithmicindent} \textbf{Output:} Optimal prospective design setting $\vb*{d}_{}^\star$
    \begin{algorithmic}[1]
    \State Initialize the design setting $\vb*{d}^{(1)}$.
        \For {$l = 1:N_\mathrm{I}$}
          \State Draw $N_\mathrm{EIG}+1$ samples $\left(\vb*{\theta}^{(0)},\vb*{\theta}^{(1)},\cdots, \vb*{\theta}^{(N_\mathrm{EIG})}\right)$ from prior $p(\vb*{\theta })$.
        \State Uniformly draw $N_\mathrm{B}$ batch samples $\left(\vb*{\theta}_\mathrm{B}^{(1)},\cdots, \vb*{\theta}_\mathrm{B}^{(N_\mathrm{B})}\right)$ from $\left(\vb*{\theta}^{(1)},\cdots, \vb*{\theta}^{(N_\mathrm{EIG})}\right)$.
        \State Evaluate the forward model $\btY_\mathrm{B}$ at the batch samples $\vb*{\theta}_{\mathrm{B}}$ and design $\vb*{d}^{(l)}$.
        \For {$i = 1:N_\mathrm{EIG}$}
         \State
        Approximate the model evaluation at the target sample $\vb*{\theta}^{(i)}$ using the RBF-based interpolation $\bhY^{(i)}\leftarrow\mathrm{RBF}$($\vb*{\theta}_\mathrm{B}$, $\btY_\mathrm{B}$, $\vb*{\theta}^{(i)}$; $n$, $m$, $q$).
        \State Draw the observation $\vb*{Y}^{(i)}$ around $\bhY^{(i)}$ using the likelihood function $p(\vb*{Y}^{(i)}\vert\bhY^{(i)})$.
        \EndFor
        \State Evaluate the EIG for $\vb*{d}^{(l)}$ using the NMC estimator in \cref{NMC}.
        \State  Obtain next search point $\vb*{d}_{}^{(l+1)}$ via random search or Bayesian Optimization (optional).
        \EndFor
        \State Find the optimal design $\vb*{d}_{}^\star\leftarrow \argmax_{1\leq l\leq N_{\mathrm{I}}}{\{\mathrm{EIG}(\vb*{d}_{}^{(l)}) \}}$.
    \end{algorithmic} 
    \begin{algorithmic}[1]
        \Function{RBF}{$\vb*{\phi}_\mathrm{B}$, $\btY_\mathrm{B}$, $\vb*{\phi}^{(i)}$; $n$, $m$, $q$} 
            \State Find the nearest $n$ batch samples $\left(\vb*{\phi}_\mathrm{B}^{(i_1)},\cdots, \vb*{\phi}_\mathrm{B}^{(i_n)}\right)$ around $\vb*{\phi}^{(i)}$ using \texttt{knnsearch}.
            \State Solve the RBF interpolation weights $\vb*{w}^{(i)}$ through \cref{rbf_poly_compact} using the RBF parameters $(m,q)$.
            \State Obtain the approximate of model evaluation as $\bhY^{(i)}\leftarrow  \sum_{j=1}^n w^{(i)}_{j} \btY(\vb*{\phi}_\mathrm{B}^{(i_j)})$. 
        \EndFunction
    \end{algorithmic}
\end{algorithm}

\subsection{Error analysis for EIG}

Next, we perform a rigorous error analysis for RBF--BOED in the estimation of the EIG.
We begin by defining the surrogate-based likelihood function and the corresponding evidence as
\begin{align}
    \widehat{p}(\vb*{Y}\vert\vb*{\theta}, \vb*{d}) 
     &\sim \mathcal{N}(\bhY(\vb*{\theta},\vb*{d}),\widehat{\vb*{\Sigma}}),\label{likelihood_interp} \\
    \widehat p(\vb*{Y}\vert \vb*{d}) & = \int_{\Theta} \widehat p(\vb*{Y}\vert\vb*{\theta}, \vb*{d}) p(\vb*{\theta}) \,\dd {\vb*{\theta}}.
\end{align}
Assuming that the interpolation outputs span the same observational space, $\mathcal{Y}$, 
we define the interpolated-based EIG analogously to \cref{EIG2} as
\begin{align}\label{EIG_interp}
     \widehat{\mathrm{EIG}}(\vb*{d})\equiv  &~\mathbb{E}_{\Theta} \{\mathbb{E}_{\mathcal{Y}\vert{\Theta,\vb*{d}}}\{  \log{ {\widehat p( \vb*{Y}\vert \vb*{\theta}, \vb*{d})}-\log {\widehat p(\vb*{Y}\vert\vb*{d})}}  \}  \}.
\end{align}
To assess the error introduced by RBF approximation, we consider the difference,
\begin{align}
    \delta \equiv \mathrm{EIG}(\vb*{d})-\widehat{\mathrm{EIG}}(\vb*{d}),
\end{align}
which quantifies the discrepancy between the true and approximated EIG.
\citet{foster2019variational} proved that \cref{EIG_interp} provides a Donsker--Varadhan lower bound of the true EIG~\citep{donsker1975asymptotic}, i.e., $\delta\geq0$, if the approximated evidence is computed consistently from the approximated likelihood.
Given the positivity of any KL divergence, an upper bound of $\delta$ can be obtained as
\begin{align}
 \delta
 &= \mathbb{E}_{ \Theta} \left[\infdiv{p(\vb*{Y}\vert\vb*{\theta},\vb*{d})}{\widehat{p}(\vb*{Y}\vert\vb*{\theta},\vb*{d})} \right]
 -\infdiv{p(\vb*{Y}\vert\vb*{d})}{\widehat{p}(\vb*{Y}\vert\vb*{d})} \\
 &\leq \mathbb{E}_{ \Theta} \left[\infdiv{p(\vb*{Y}\vert\vb*{\theta},\vb*{d})}{\widehat{p}(\vb*{Y}\vert\vb*{\theta},\vb*{d})} \right].\label{delta_UB1}
\end{align}
This upper bound is closely related to the established Barber--Agakov variational bound, 
which similarly uses conditional KL divergence to approximate mutual information~\citep{barber2004algorithm}.
Therefore, the error satisfies $\delta\in\left[0, \mathbb{E}_{ \Theta} \left[\infdiv{p(\vb*{Y}\vert\vb*{\theta},\vb*{d})}{\widehat{p}(\vb*{Y}\vert\vb*{\theta},\vb*{d})} \right] \right]$ and the bounds are tight if and only if $\widehat{p}(\vb*{Y}\vert\vb*{\theta}, \vb*{d}) ={p}(\vb*{Y}\vert\vb*{\theta}, \vb*{d})$, or $\bhY=\btY$, for almost all $\vb*{\theta}$.
In the special case where both the true and interpolation-based likelihoods are Gaussian, the upper bound in \cref{delta_UB1} simplifies to
\begin{align}
 \delta
 \leq \frac{1}{2}\mathbb{E}_{ \Theta}
 \left[\mathrm{tr}(\widehat{\vb*{\Sigma}}^{-1}\vb*{\Sigma})+
\underbrace{(\btY -\bhY ){\widehat{\vb*{\Sigma}}_{}^{-1}} (\btY -\bhY )^\top}_{\delta_M^2}
 -N_t+\log\left(\frac{\det \widehat{\vb*{\Sigma}}}{\det \vb*{\Sigma}}\right)
 \right], \label{delta_UB2}
\end{align}
where $\delta_M$ denotes the Mahalanobis distance between $\btY$ and $\bhY$ under the covariance matrix $\widehat{\vb*{\Sigma}}$.
While variational Bayesian and implicit likelihood approaches also use surrogate EIGs, our method fundamentally differs by approximating deterministic model outputs rather than introducing stochastic generative surrogates.
Equation \Cref{delta_UB2} quantifies how RBF interpolation errors propagate to the EIG estimates.
For example, an interpolation with second-order accuracy has $\delta_M = O\left((\delta \vb*{\theta})^2\right)$, which leads to an EIG estimate with fourth-order accuracy, 
$\delta = O\left(\mathbb{E}_{\Theta}\{(\delta \vb*{\theta})^4\}\right)$,
where $\delta \vb*{\theta}$ is the local distances between samples.

\subsection{Validation: 3D nonlinear design}

To demonstrate the performance and assess the error of the RBF--BOED approach, 
we adopt the nonlinear model introduced by \citet{dong2025variational} as a benchmark test problem. 
The nonlinear observation model is defined as
\begin{align*}
    y = \tilde{y}(\vb*{\theta},d)+\epsilon, \quad \text{where} \quad \tilde{y}(\vb*{\theta},d) = \theta_1^3 d^2+\theta_2 \mathrm{exp}\left(-|0.2-d|\right) +\sqrt{2\theta_3^2d},
\end{align*}
with independent prior on parameters $\theta_1\sim \mathcal{N}(0.5,0.3^2)$
, $\theta_2\sim \mathcal{N}(0.3,0.7^2)$, $\theta_3\sim \mathcal{N}(0.5,0.8^2)$.
A univariate design variable, $d$, is considered within the interval $d\in[0,1]$.
The measurement noise follows a Gaussian mixture distribution, $\epsilon\sim \mathcal{N}(0.1,0.05^2)/2+\mathcal{N}(-0.1,0.05^2)/2$.
Since only a single forward model is introduced in this case, the parameter $\vb*{\theta}$ is equivalent to the parameter $\vb*{\phi}$ defined in \cref{phi_M}.

\begin{figure}[ht!]
    \centering
    \includegraphics[scale=1]{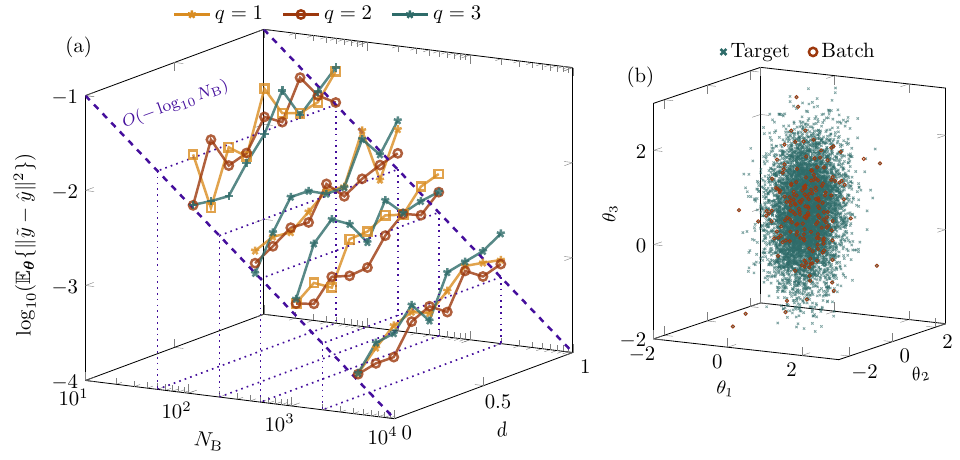}
    \caption{
        Error analysis of the RBF--BOED method across varying batch sample sizes: $N_\mathrm{B}=50,\,200,\,500,\,1000.$
    }
    \label{f:toy_error}
\end{figure}

The RBF--BOED approach leads to the interpolation within the given 3D parameter space.
\Cref{f:toy_error}~(a) presents the corresponding error across various batch sample sizes.
As an example, \cref{f:toy_error}~(b) shows a set of randomly selected batch samples with a size of $N_\mathrm{B}=200$.
For augmented polynomials of different orders, the ensemble-averaged squared distance shows a consistent slope of approximately $-1$, suggesting a convergence rate of the RBF interpolation on the order of $O(N_\mathrm{B}^{-2})$.
To minimize variance while maintaining low error within the design space, we select a polynomial augmentation order of $q=2$.
This choice balances accuracy and stability, avoiding overfitting risks associated with higher-order terms.

\begin{figure}[ht]
    \centering
    \includegraphics[scale=1]{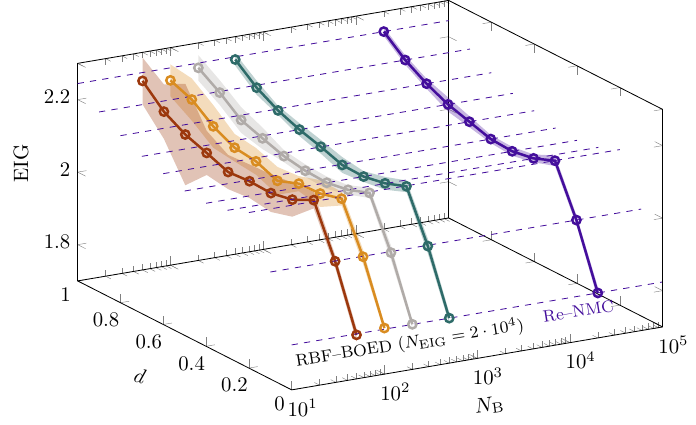}
    \caption{
Comparison of ensemble-averaged EIG estimates using the RBF--BOED method across varying batch sample sizes.
The Re--NMC estimator with $N_{\mathrm{EIG}}=2\cdot10^4$ serves as a baseline reference. 
Shaded uncertainty regions show $\pm 2$ standard deviations.
    }
    \label{f:toy_EIG}
\end{figure}

\Cref{f:toy_EIG} compares the EIG estimates from the RBF--BOED method against the benchmark Re--NMC estimator ($N_{\mathrm{EIG}}=2\cdot10^4$).
As demonstrated by \citet{dong2025variational}, the Re--NMC estimator achieves near-identical accuracy to the high-fidelity DLMC estimator while requiring $4\cdot10^8$ forward simulations.
In contrast, the RBF--BOED method reduces the number of forward simulations by orders of magnitude compared to Re--NMC.
As expected, the RBF--BOED provides a lower bound of the EIG when the added noise follows a constant distribution. 
Even for a batch sample size of $N_{\mathrm{B}}=50$, RBF--BOED produces qualitatively accurate EIG estimates.
The variability of EIG estimates diminishes as $N_{\mathrm{B}}$ increases.
At $N_{\mathrm{B}}=500$, the results converge to those achieved with $N_{\mathrm{B}}=2\cdot10^4$, resulting in a $2.5\%$ computational cost.
For $N_{\mathrm{B}}=200$ ($1\%$ of the cost), 
the estimates exhibit slightly higher variance but retain accuracy in mean EIG values.
The computational advantage of RBF--BOED becomes particularly significant when applied to expensive forward models.

\begin{figure}[ht]
    \centering
    \includegraphics[scale=1]{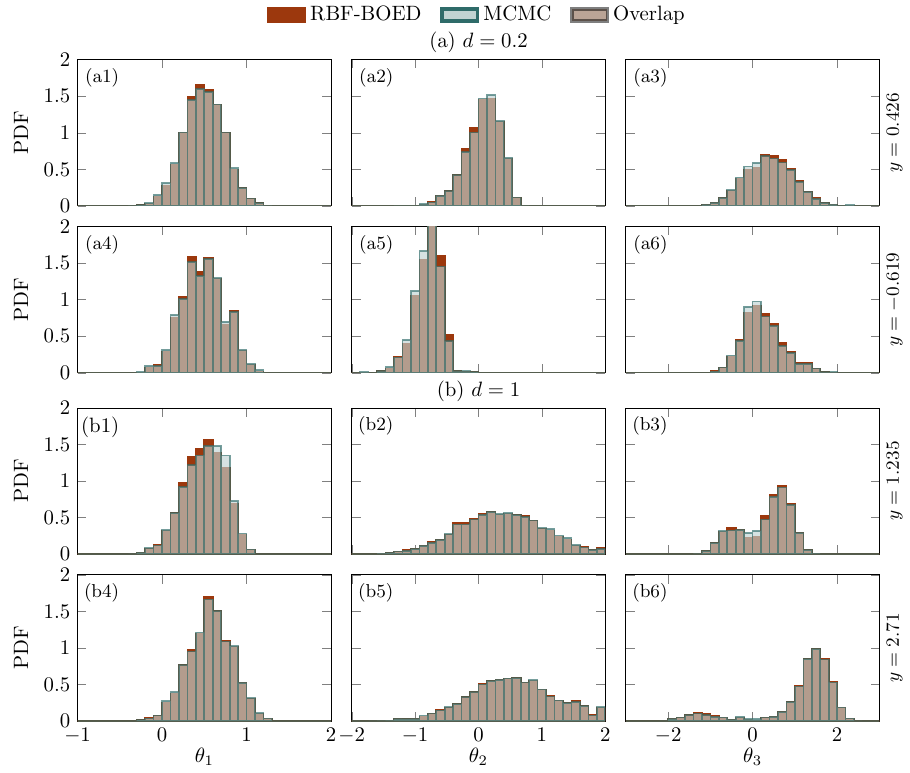}
    \caption{
        Marginal posteriors, $p(\vb*{\theta}\vert y,d)$, obtained using RBF--BOED ($N_{\mathrm{B}}=200$) and MCMC ($N_{\mathrm{EIG}}=2\cdot10^4$): (a) $d=0.2$; (b) $d=1$.
    }
    \label{f:toy_posterior}
\end{figure}

To assess RBF--BOED’s capability in approximating true posterior distributions, we compare posteriors derived using $N_\mathrm{B}=200$ to those generated via Markov chain Monte Carlo (MCMC) using $N_\mathrm{EIG}=2\cdot10^4$ simulations, a standard benchmark~\citep{andrieu2003introduction}.
\Cref{f:toy_posterior} shows the marginal posteriors at experimental designs and observational conditions matching those in \citet{dong2025variational}.
Despite using only $1\%$ of the total forward simulations, RBF--BOED closely matches MCMC posteriors while preserving key features of the ground-truth distributions.
Notably, it accurately represents the multi-modal posteriors in \cref{f:toy_posterior}~(b3,\,b6), which are particularly challenging to approximate due to sampling inefficiency.
These results highlight RBF--BOED’s effectiveness in balancing statistical fidelity with computational efficiency.

\section{Applications: optimal LIC experimental designs} \label{S:results}

We apply the RBF--BOED approach in two distinct design scenarios for LIC experiments, distinguished by differing levels of model complexity.
The first scenario employs a single constitutive model, while the second incorporates multiple constitutive models with equal prior probabilities.
The prior distributions, shown in \cref{datasets}, are parameterized to align with experimental data from \citet{yang2020extracting}, ensuring consistency with real-world material behavior. 
Computations are conducted on PSC~Bridges2 using dual AMD 64-core CPUs (SKU~7742, Rome). 
The computational cost differs across models due to varying automatic time step requirements for handling stiffness near bubble collapse.
On average, generating $N_{\mathrm{B}}=10^4$ IMR simulations for the multi-model design case requires approximately 300~CPU~core~hours.
To emulate experimental observations, time- and state-dependent synthetic noise is added to the IMR simulations, with standard deviations of $\sigma = |R^*-1|/20+(1-\exp( -t^*/10))/30$.
This formulation qualitatively reflects noise characteristics in previous experiments~\citep{estrada2018high, yang2020extracting, yang2022mechanical}.
Bubble radius measurements are sampled over $t^*\in\left[0,3\right]$ with a constant time step of $\mathrm{d}t^*=0.05$, yielding an observation vector of dimension $N_t = 61$.
In the following, we demonstrate how the RBF--BOED approach achieves significant computational savings in the LIC experimental designs while maintaining robustness to model complexity and noise.
All parameters are normalized and standardized.

\begin{table}[ht!]
    \centering
    \caption{
        Summary of parameter priors for each model as indicated.
        (---) indicates inapplicable parameters.
    }
    \label{datasets}
    \begin{tabular}{c c c c c c c c}
          \multirow[b]{2}{*}{Model $\mM$} 
        &  \multicolumn{5}{c}{Parameters $\vb*{\phi}_{\mM}$} \\
    &   $G_{\infty}$ [\unit{\kilo\pascal}] 	 & $G$ [\unit{\kilo\pascal}] & $G_1$ [\unit{\kilo\pascal}]  & $\mu$ [\unit{\pascal\,\second}] & $\alpha$   \\\midrule
    LKV & --- & $15.09\pm4.35$ & ---     &$0.209\pm0.10$  & --- \\ 
    NHKV & --- & $15.09\pm4.35$ & --- & $0.209\pm0.10$     & ---  \\
     Gen.\ qKV & $2.77\pm 0.3$ &  ---   & ---  & $0.286\pm0.14$ &  $0.28\pm0.48$  \\
      SLS & --- & $15.09\pm4.35$ & $10^{6\pm2}$     &$0.209\pm0.10$  & --- \\ 
     SNS & --- & $15.09\pm4.35$ & $10^{6\pm2}$    &$0.209\pm0.10$  & --- 
    \end{tabular}
\end{table}

\subsection{Case 1: Single-constitutive-model design}

\begin{figure}
    \centering
    \includegraphics[scale=1]{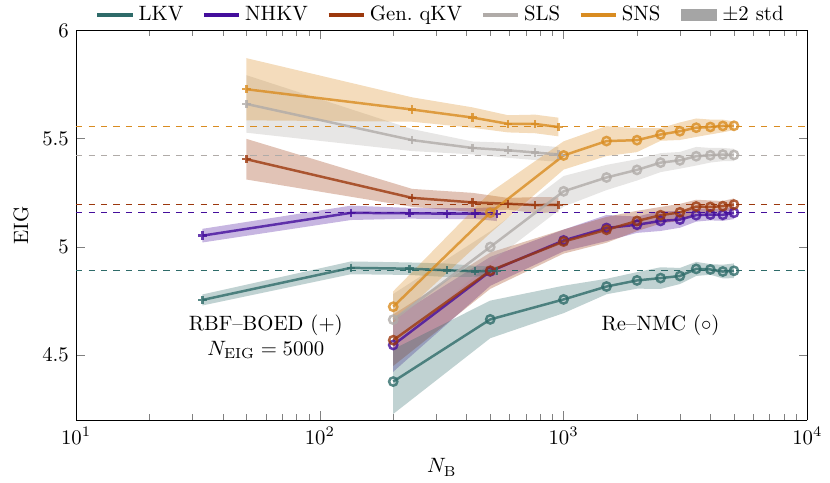}
    \caption{
   Comparison of the RBF--BOED and Re--NMC methods in single-constitutive-model LIC experiments at $\vb*{d}=(\mathrm{We},R^*_{\infty})=(833,0.28)$.
   Ensemble-averaged EIG estimates are presented, with shaded regions representing $\pm 2$ standard deviations.
    }
    \label{f:EIG_IMR_1model}
\end{figure}

We begin by considering the simplest case, in which a single constitutive model is presumed to represent the experimental data accurately.
This baseline isolates the impact of measurement noise from uncertainties associated with model selection.
\Cref{f:EIG_IMR_1model} compares the EIG estimates obtained using the RBF--BOED and Re--NMC methods for the design $\vb*{d}=(\mathrm{We},R^*_{\infty})=(833,0.28)$.
At this setting, the Re--NMC estimator achieves convergence around a sample size of $N_\mathrm{EIG}=5000$ across all five models. 
We therefore adopt it as a reference for evaluating the performance of the RBF--BOED approach.
For the LKV and NHKV models, each involving two free material parameters, a batch sample size of fewer than $N_{\mathrm{B}} = 200$ suffices to recover the high-fidelity EIG accurately, yielding a computational cost reduction of approximately $96\%$.
As expected, the more complex three-dimensional constitutive models—Gen.\ qKV, SLS, and SNS—require a greater number of IMR simulations to achieve similar levels of convergence.
When the batch sample size is too small ($N_\mathrm{B}\lesssim 200$), the observation space $\mathcal{Y}$ containing $N_t=61$ time steps is not adequately represented by the available samples within the 3D parameter space, resulting in extrapolations rather than reliable interpolations in sparsely sampled regions.
Due to inaccurate surrogate predictions in sparsely sampled regions, higher variability and overestimated EIG are observed.
As the batch sample size increases, the transition from error-prone extrapolation to robust interpolation improves the approximation of $\mathcal{Y}$ and reduces the overprediction error.
Allocating approximately $10\%$ of the full computational cost ($N_\mathrm{B} \approx 500$) is sufficient to produce reasonably accurate EIG estimates, demonstrating the adaptability of RBF--BOED in high-dimensional parameter spaces.

\begin{figure}[ht!]
    \centering
    \includegraphics[scale=1]{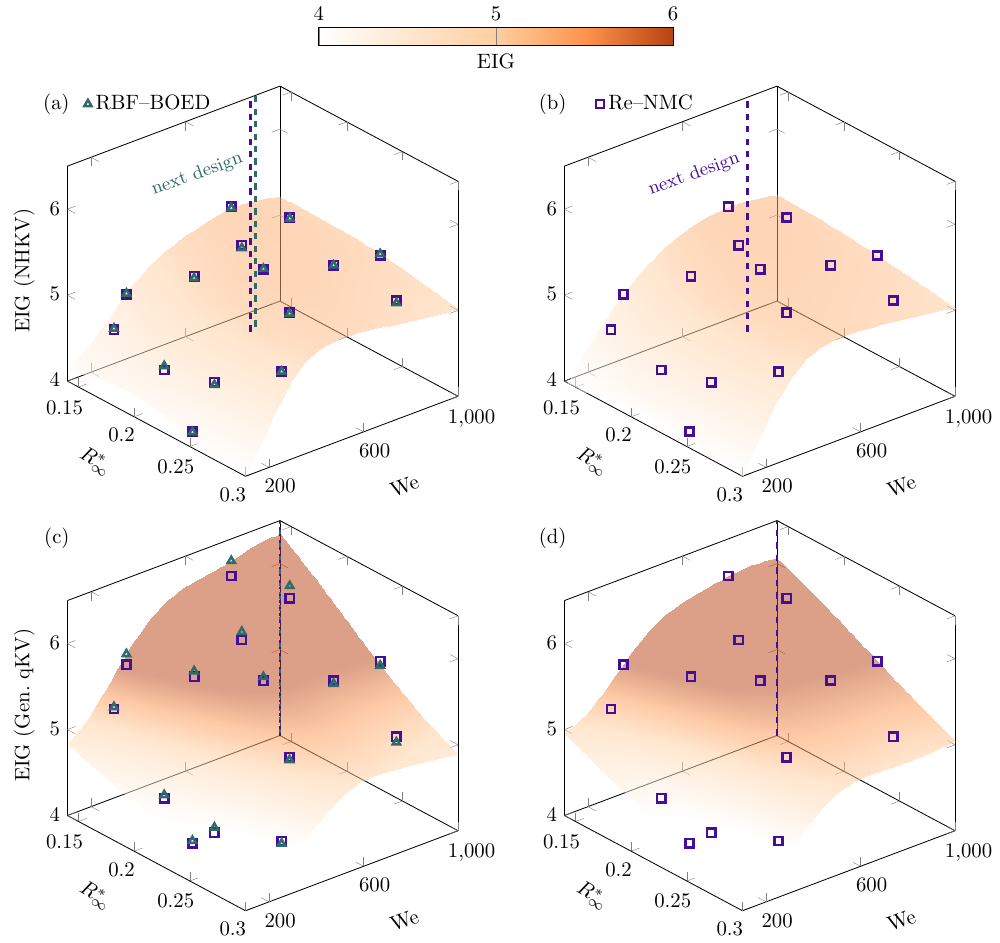}
    \caption{
        EIG evaluations and corresponding GP regressions for single-constitutive-model LIC experiments: (a, b) NHKV; (c, d) Gen.\ qKV.
        RBF--BOED is applied with $N_{\mathrm{B}}=200$ for NHKV (a) and $N_{\mathrm{B}}=500$ for Gen.\ qKV (c), whereas the baseline Re--NMC approach uses $N_{\mathrm{EIG}} = 5000$ simulations in both cases (b, d). 
        Using the same BO process, RBF--BOED effectively identifies the next experimental design (dashed lines) using less than $10\%$ of the computational cost required by Re--NMC.
    }
    \label{f:EIG_IMR_1model_fields}
\end{figure}

The RBF--BOED enables rapid EIG evaluation, therefore enabling efficient design space exploration via BO.
\Cref{f:EIG_IMR_1model_fields} shows 15 EIG evaluations randomly sampled from the design space of the single-constitutive-model LIC experiments and their corresponding GP regressions obtained using~\cref{M52}.
Within the given design space, the EIG exhibits a minimum value exceeding 4, reflecting a baseline level of informativeness inherent to the LIC experiments.
We highlight that the objective of BOED is to identify the design that maximizes EIG, regardless of its absolute magnitude.
Compared to the baseline Re--NMC method, RBF--BOED achieves some percentage of closeness of one another in EIG estimations across various design points for both models under study, while reducing computational costs to $4\%$ for NHKV and 10$\%$ for Gen.\ qKV.
For both models, increasing the Weber number results in greater EIG, suggesting that inertial-dominated regimes yield more informative experimental outcomes.
Furthermore, the EIG of the Gen.\ qKV model exhibits a stronger dependence on the dimensionless equilibrium radius, $R_{\infty}^*$, compared to the NHKV model.
This contrast underscores how distinct physical mechanisms—such as viscoelastic relaxation versus inertial effects—drive information acquisition in each constitutive model.
By maintaining high-fidelity GP surrogates at reduced computational cost, RBF--BOED enables efficient design-space exploration and accurate selection of subsequent designs.
At the same time, the method preserves the interpretability of the underlying physical mechanisms.


\subsection{Case 2: Multi-constitutive-model design}

Next, we address the practical scenario where the constitutive model governing the material response is unknown \textit{a priori}, a common challenge in rheology.
Here, we adopt a Bayesian model-averaging approach, assigning equal prior probabilities to all five candidate models.
This model-agnostic approach enables the optimization of LIC designs while explicitly accounting for structural uncertainty.
The RBF--BOED method consistently achieves a reduced computational cost compared to baselines: $4\%$ for 2D and $10\%$ for 3D constitutive models.

\begin{figure}
    \centering
    \includegraphics[scale=1]{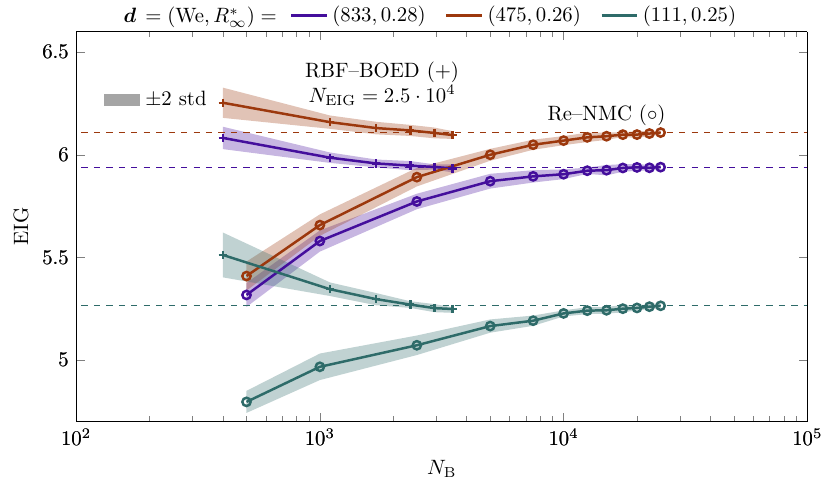}
    \caption{
        Ensemble-averaged EIG estimates for the 5-constitutive-model LIC experiments across varying design settings and batch sample sizes.
        Shaded uncertainty regions show $\pm 2$ standard deviations.
    }
    \label{f:EIG_IMR_5models}
\end{figure}

\Cref{f:EIG_IMR_5models} shows the EIG estimates for the 5-model LIC experiments across varying design settings and batch sample sizes.
The baseline Re--NMC estimator requires $N_{\mathrm{EIG}}=2.5\cdot10^4$ samples to achieve convergence, whereas the RBF--BOED method attains comparable accuracy with fewer than $N_{\mathrm{B}}=2000$ batch IMR simulations.
This results in a computational cost reduction exceeding 92$\%$ while maintaining accuracy.
As discussed in the context of \cref{f:EIG_IMR_1model}, 
insufficient sample size ($N_\mathrm{B}\lesssim 1000$) fails to adequately capture the 61-dimensional observation space $\mathcal{Y}$, particularly within the three-dimensional parameter spaces, leading to an overestimated EIG.
Increasing $N_\mathrm{B}$ to 2000 enables robust EIG quantification across diverse design configurations, 
demonstrating that the information about the material properties in the 5-constitutive-model system is reliably captured.

\begin{figure}[ht!]
    \centering
    \includegraphics[scale=1]{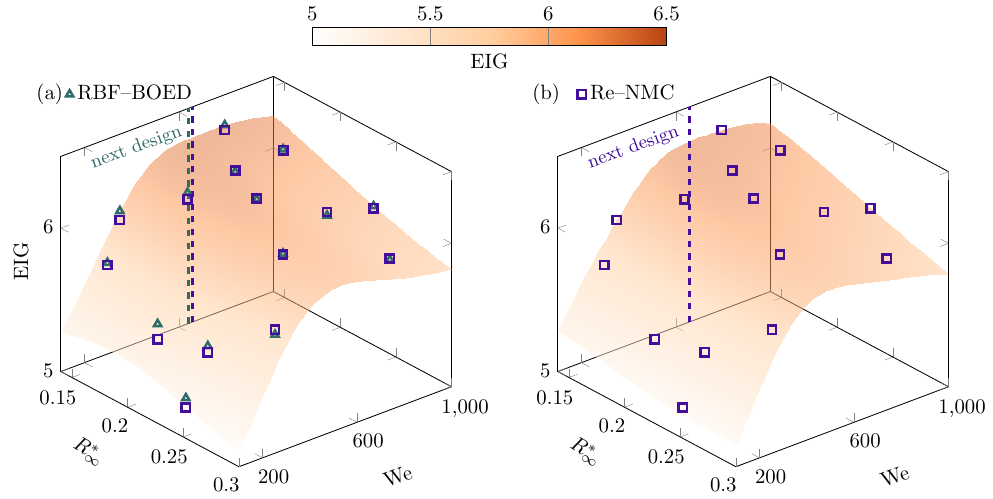}
    \caption{
EIG evaluations and the corresponding GP regressions for the 5-constitutive-model LIC experiments: (a) RBF--BOED with $N_{\mathrm{B}}=2000$ and $N_{\mathrm{EIG}}=2.5\cdot10^4$; (b) Re--NMC with $N_{\mathrm{B}}=N_{\mathrm{EIG}}=2.5\cdot10^4$.
Through the same BO process, RBF--BOED effectively identifies the next experimental design (dashed lines) using $8\%$ of the computational cost required by Re--NMC.
    }
    \label{f:EIG_IMR_5models_fields}
\end{figure}

\Cref{f:EIG_IMR_5models_fields} compares the EIG landscapes generated via the RBF--BOED and the baseline Re--NMC methods, based on 15 random evaluations each.
The RBF--BOED approach accurately identifies EIG estimates across the design space and their corresponding GP regression.
The computational cost is reduced to $8\%$ of the baseline requirement that uses $N_{\mathrm{EIG}}=2.5\cdot10^4$ simulations.
Although the EIG landscape of the five-constitutive-model system deviates from those of the NHKV and Gen.\ qKV models in \cref{f:EIG_IMR_1model_fields}, low Weber numbers consistently correlate with reduced EIG values across all configurations, reflecting decreased information acquisition in viscous-dominant regimes where inertial effects are secondary.
We identify an optimal EIG region of low equilibrium radius ($R_{\infty}^*\lesssim 0.2$) and high Weber number ($\mathrm{We}\gtrsim 500$) to guide subsequent experimental design.
This information can be acquired via the RBF--BOED method using just 900~CPU~core~hours, bypassing the need for high-fidelity EIG computations that require 11250~CPU~core~hours.
Therefore, a 12.5-fold acceleration is achieved in the BOED process for LIC experiments.

\section{Limitations of present work}\label{s:limits}

Local RBF methods offer computational efficiency for low-dimensional interpolation tasks, such as in two- or three-dimensional parameter spaces.
However, they face challenges in high-dimensional parameter spaces, particularly when coupled with high-dimensional observation spaces.
In such scenarios, constructing accurate local surrogates becomes increasingly difficult, as the number of required sample points grows rapidly with dimensionality, potentially degrading surrogate quality and increasing computational overhead.
Incorporating dimensionality reduction techniques or adaptive stencil selection strategies may help mitigate these issues and improve scalability.

The current version of RBF--BOED approach requires knowledge of the likelihood function as a prior.
However, this is not a limitation when accurate physical models are available to represent the observations: the Keller--Miksis equations describe spherical bubble dynamics in this application.
However, the applicability of RBF--BOED is limited when the likelihood is intractable or only implicitly defined, such as in likelihood-free inference or models with complex simulator-based outputs.
This limitation may be addressed by incorporating strategies analogous to those used in GP models, where variance structures are introduced alongside deterministic global RBF kernels to account for stochasticity.
Likewise, local RBF methods can be extended to incorporate local variance estimates, enabling compatibility with likelihood-free methods that rely solely on sampled likelihood values.
This adaptation would broaden the scope of RBF--BOED, making it suitable for a wider class of models encountered in experimental design problems.

Additionally, the current implementation focuses on estimating the EIG field rather than its gradient and thus relies on subsequent space exploration strategies, such as BO, to identify optimal designs.
Since gradients with respect to material properties are not explicitly available, a promising direction for future work is to compute Jacobians of forward simulations from existing batch samples using RBF-based finite differences (RBF--FD)~\citep{fornberg2011stabilization,flyer2016enhancing,flyer2016role}.
These Jacobians could then be used to estimate the gradient of the EIG, enabling the integration of gradient-based optimization methods.
This adaptation would improve optimization efficiency and provide a more direct route to identifying optimal experimental designs, particularly in high-dimensional parameter spaces.
Application-specific challenges remain, however.
For bubble dynamics trajectories, the inherent rapid changes during the collapse and rebound phases pose significant challenges for stable gradient estimation.
Temporal segmentation of trajectories may be helpful to address these discontinuities while preserving numerical accuracy.

\section{Conclusions}\label{s:conclusions}

This study develops an efficient Bayesian optimal experimental design (BOED) approach based on the local radial basis functions (RBFs).
The capability of RBFs to handle scattered data aligns naturally with probabilistic sampling.
The localized stencil structure further leverages the smoothness of forward model responses to material properties, enabling high-order approximations while maintaining low computational costs.
By integrating these features, the method constructs accurate and robust deterministic surrogates.
The RBF-BOED approach reduces reliance on costly forward model evaluations by strategically sampling a sparse subset of prior parameters, enabling efficient estimation of expected information gain (EIG).
This combination of computational efficiency, accuracy, and robustness positions the RBF--BOED method as a promising tool for complex and resource-constrained design problems.

To demonstrate its practical utility, we apply the RBF--BOED method to optimize the laser-induced cavitation (LIC) experimental setup.
Forward simulations are performed using inertial microcavitation rheometry (IMR), a technique for characterizing the viscoelastic properties of hydrogels by analyzing the time history of bubble radius evolution.
Two design scenarios of increasing complexity—single-model and multi-model LIC design problems—are examined to illustrate the method’s performance and generality.
Notably, for a five-constitutive-model design problem, the RBF--BOED approach achieves accurate EIG estimation at only $8\%$ of the full computational cost, demonstrating its efficiency and adaptability for complex experimental designs.

\section*{Data availability}

The codes are available at \url{https://github.com/InertialMicrocavitationRheometry/IMR_RBF_BOED}.

\section*{CRediT authorship contribution statement}

\textbf{TC}: Formal analysis, Methodology, Software, Investigation, Data Curation, Validation, Visualization, Writing – original draft, Writing – review $\&$
editing.
\textbf{JBE}: Conceptualization, Funding acquisition, Methodology, Project administration, Resources, Writing – review $\&$
editing.
\textbf{SHB}: Conceptualization, Funding acquisition, Methodology, Project administration, Resources, Supervision, Writing – review $\&$
editing.

\section*{Declaration of competing interest}

The authors declare that they have no known competing financial interests or personal relationships that could have appeared to influence the work reported in this paper.

\section*{Acknowledgments}

The authors acknowledge support from the U.S.\ Department of Defense, the Army Research Office under Grant No. W911NF-23-10324 (PMs Drs.\ Denise Ford and Robert Martin).
This work used PSC~Bridges2 and NCSA~Delta through allocation PHY210084 (PI Bryngelson) from the Advanced Cyberinfrastructure Coordination Ecosystem: Services $\&$ Support (ACCESS) program~\citep{boerner2023access}, which is supported by National Science Foundation grants $\#$2138259, $\#$2138286, $\#$2138307, $\#$2137603, and $\#$2138296.
T.C.\ thanks Jiayuan Dong for generously sharing code related to variational Bayesian optimal experimental design using normalizing flows.

\bibliographystyle{bibsty}
\bibliography{references}

\end{document}